\documentclass[amsmath,amssymb]{revtex4}
\usepackage{graphicx}% Include figure files
\usepackage{amssymb}
\usepackage{amsmath}
\usepackage{color}

\begin{document}

%\draft
\title{Non-Markovian dynamics with  fermions}
\author{
V.V. Sargsyan$^{1}$, G.G. Adamian$^{1}$,  N.V. Antonenko$^{1}$, and D. Lacroix$^{2}$
}
\affiliation{
$^{1}$Joint Institute for Nuclear Research, 141980 Dubna, Russia\\
$^{2}$Institut de Physique Nucl\'eaire, IN2P3-CNRS, Universit\'e Paris-Sud, F-91406 Orsay Cedex, France
}
\date{\today}

\begin{abstract}
Employing the quadratic fermionic Hamiltonians for the collective and internal subsystems with a linear coupling, we studied
the role of fermionic statistics on the dynamics of the collective motion.
The transport coefficients are discussed as well as the associated fluctuation-dissipation relation.
Due to  different nature of the particles, the path to equilibrium
is slightly affected.
However,  in the weak coupling regime, the time-scale for approaching equilibrium is found to be globally unchanged.
The Pauli-blocking effect can modify the usual picture in open quantum system.
In some limits, contrary to boson, this effect can strongly hinder
the influence of the bath by blocking the interacting channels.
\end{abstract}


 \maketitle

\section{Introduction}
Investigations of the behavior of dissipative quantum non-Markovian collective subsystems beyond the weak coupling
or high temperature limits triggered significant interest in exactly solvable
models \cite{M0,M1,M2,M4,M5,M6,M7,M8,M10,M11,M12,M13,M14,M15}.
The internal subsystem in these models is represented by a collection of
bosonic harmonic oscillators, which interacts with the collective bosonic subsystem,
modeled by a harmonic oscillator as well, via a linear coupling
between the coordinates and/or momenta. The quantum-mechanical
motion of a fermionic oscillator (two-level collective subsystem)
coupled linearly to dissipative fermionic environment
requires an intensive study \cite{M6}.
The interest in considering fermionic baths is however growing up due to the possibility of creating and manipulating
 rather small fermionic systems in condensed matter, atomic and nuclear physics. In particular, the approaching phase to equilibrium of these systems
would greatly help to understand how they can reach the thermodynamic limits \cite{Rig14}.

In the present paper, we  use the quadratic fermionic Hamiltonians for the collective and internal subsystems with a linear coupling, and present
the detailed analysis of the role of the fermionic statistics (in the comparison with the bosonic one)  in the dynamics of the collective motion.
As an example, we compare the decays of excited collective states of the bosonic and fermionic harmonic oscillators.

We use the Langevin approach \cite{M11,M12,M13,M14,M15}
which is widely applied to find the effects of fluctuations and dissipations in macroscopical systems.
The Langevin method in the kinetic theory significantly simplifies the calculation of non-equilibrium
quantum and thermal fluctuations and provides a clear picture of the dynamics.
Many problems in solid state physics, condensed matter, chemistry, nuclear and atomic physics can be described using the
Langevin equations in the space of relevant collective coordinates.

In Sec. II.A we
present a fully quantum-mechanical derivation of   non-Markovian Langevin equations. These equations
fulfill the quantum fluctuation-dissipation theorem as shown
in Sec. II.B.  In Secs. II.C and II.D the analytical expressions for occupation number are derived.
The time-dependent transport coefficients which
take the memory effects into consideration are obtained in Sec. II.E.   The results of illustrative numerical calculations of diffusion and friction coefficients and
level populations are presented in Sec. III.
Sec. III  deals with the decay  of excited state for the collective plus internal bosonic  or  fermionic subsystems.
A summary is given in Sec. IV.

\section{Formalism}

\begin{figure}[h]
\includegraphics[scale=0.5]{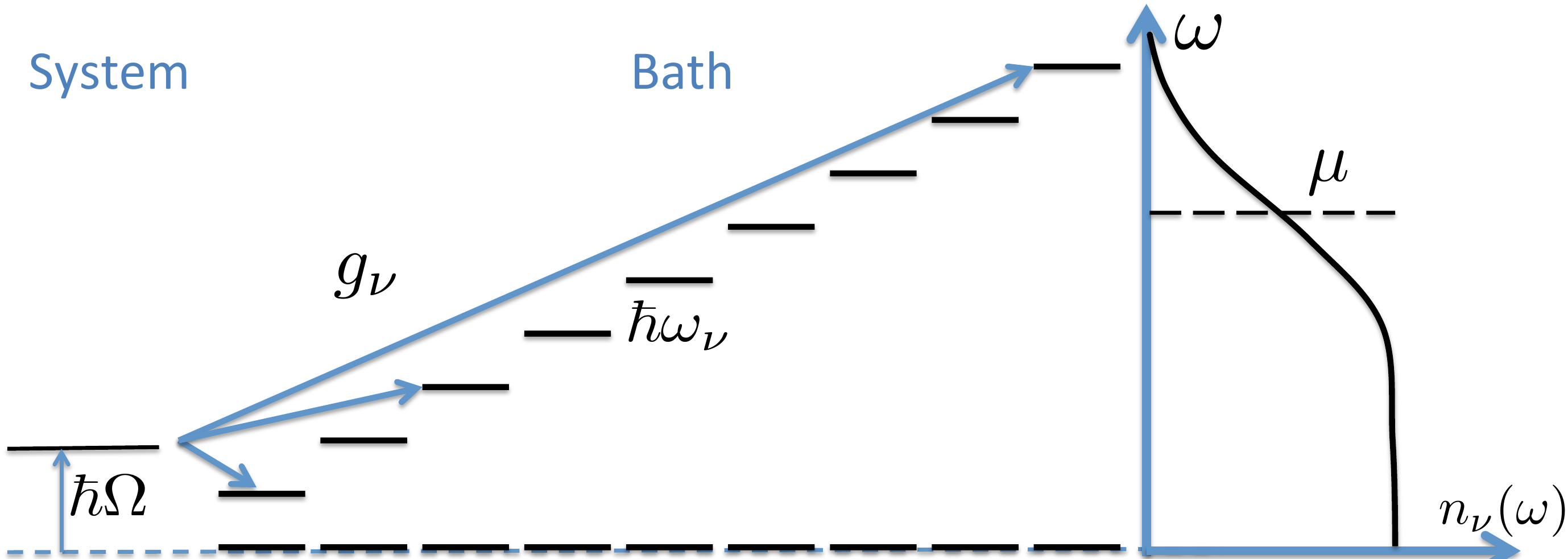}
\caption{
(Color online) Schematic representation of the system coupled to the bath of two-level fermionic systems.
The systems is assumed to be only coupled to the excited states in each two-level. $\mu$ is the chemical
potential that, at a given temperature, can be used to modify the average number of particles in the bath coupled to the system.
$\Omega$ is the renormalized frequency of the system (see text).
In the right side, a typical occupation probability profile for the bath is shown.
}
\label{figure0}
\end{figure}

The Hamiltonian of whole system (the heat bath plus collective subsystem) is assumed here to have the following
form  \cite{M6}:
\begin{equation}
H=H_c+H_b+H_{cb}=\hbar \omega a^ + a+ \sum_{\nu } \hbar \omega _{\nu }a_{\nu }^+a_{\nu } + \sum _{\nu } g_{\nu }\left(a_{\nu }^+a+a^+a_{\nu }\right),
\label{eq_ham}
\end{equation}
which explicitly depends on the collective (heat bath)
creation $a^{+}$ ($a_{\nu }^{+}$) and annihilation $a$ ($a_{\nu }$) operators. Contrary to the usual assumption,
these operators are   considered here as the fermionic operators
and satisfy the following permutation relations:
\begin{equation}
a a^++a^+a=1,  aa=a^+a^+=0, \\
a_{\nu}a_{\nu' }^++a_{\nu'}^+a_{\nu}=\delta_{\nu,\nu'},   a_{\nu}a_{\nu'}+a_{\nu'}a_{\nu}=a_{\nu}^+a_{\nu'}^++a_{\nu'}^+a_{\nu}^+=0.
\label{eq_permut}
\end{equation}
In Eq.~(\ref{eq_ham}) the terms $H_c$, $H_b$, and $H_{cb}$ are the Hamiltonians  of the collective subsystem depending on
the frequency $\hbar\omega$ of the fermionic oscillator, of the intrinsic   bath subsystem, and
of the collective-bath interaction, respectively.
The model of heat bath is an assembly of the fermionic oscillators "$\nu$"  with frequencies  $\hbar\omega_\nu$.
More precisely, we assumed that the bath is a collection of two-level fermionic systems depicted
in Fig. \ref{figure0}. The use of two-level picture is helpful to simplify  slightly  the derivations given below while keeping the main physical effects
associated with the fermionic nature of the bath.

The coupling to the heat bath is linear in the collective and bath operators and corresponds
to the energy being transferred to and from the bath or collective subsystem  by the creation and absorption
of bath or collective subsystem quanta.
The real constants $g_\nu$ determines the coupling strength between
the collective oscillator and bath oscillator "$\nu$".
The coupling term can have important consequences on the dynamics of the collective subsystem by
altering the effective collective potential and by allowing energy to be exchanged with the thermal reservoir,
thereby, allowing the collective subsystem to attain the thermal equilibrium with the heat bath.
%For simplicity we consider here
%the two-level collective and intrinsic systems.

\subsection{Equations of motion}
The  Heisenberg  equations of motion  for the creation and anihilation operators of the collective
and intrinsic subsystems are obtained by commuting corresponding operator with $H$:
\begin{eqnarray}
\frac{d}{d t}a^{+}(t)&=&i \omega a^++\frac{i}{\hbar} \sum_{\nu} g_{\nu }a_{\nu }^+, \nonumber \\
\frac{d}{d t}a(t)&=&-i \omega a -\frac{i}{\hbar }\sum_{\nu} g_{\nu }a_{\nu },
\label{eq_coll}
\end{eqnarray}
\begin{eqnarray}
\frac{d}{dt}a_{\nu }^+(t)=i \omega _{\nu } a_{\nu }^++\frac{i}{\hbar }g_{\nu }a^+, \nonumber \\
\frac{d}{dt}a_{\nu }(t)=-i \omega _{\nu } a_{\nu }-\frac{i}{\hbar }g_{\nu }a.
\label{eq_intr}
\end{eqnarray}
The solutions of Eqs.(\ref{eq_intr}) are:
\begin{eqnarray}
a_{\nu }^+(t)&=&\left(a_{\nu }^+(0)+\frac{g_{\nu }}{\hbar  \omega _{\nu }}a^+(0)\right)e^{i \omega _{\nu }t}-\frac{g_{\nu }}{\hbar  \omega _{\nu }}a^+(t)+\frac{g_{\nu }}{\hbar  \omega _{\nu }}\int _0^te^{i \omega _{\nu }(t- \tau )}\frac{da^{+}(\tau)}{d\tau}d\tau ,\nonumber \\
a_{\nu }(t)&=&\left(a_{\nu }(0)+\frac{g_{\nu }}{\hbar  \omega _{\nu }}a(0)\right)e^{-i \omega _{\nu }t}-\frac{g_{\nu }}{\hbar  \omega _{\nu }}a(t)+\frac{g_{\nu }}{\hbar  \omega _{\nu }}\int _0^te^{-i \omega _{\nu }(t- \tau )}\frac{da(\tau)}{d\tau}d\tau .
\label{eq_intr2}
\end{eqnarray}
Substituting Eqs. (\ref{eq_intr2}) in    (\ref{eq_coll}) and eliminating the bath variables from the
equations of motion of the collective subsystem, we
obtain a set of   integro-differential stochastic dissipative equations:
\begin{eqnarray}
\frac{d}{dt}a^+(t)&=&i \omega a^+ +\frac{i}{\hbar }\sum _{\nu } g_{\nu }\left(a_{\nu }{}^+(0)+\frac{g_{\nu }}{\hbar  \omega _{\nu }}a^+(0)\right)e^{i \omega _{\nu }t}-i\sum _{\nu } \frac{g_{\nu }^2}{\hbar ^2 \omega _{\nu }}a^+(t)+i\sum _{\nu } \frac{g_{\nu }^2}{\hbar ^2 \omega _{\nu }}\int _0^te^{i \omega _{\nu }(t- \tau )}\frac{da^{+}(\tau)}{d\tau}d\tau ,\nonumber \\
\frac{d}{dt}a(t)&=&-i \omega  a-\frac{i}{\hbar }\sum _{\nu } g_{\nu }\left(a_{\nu }(0)+\frac{g_{\nu }}{\hbar  \omega _{\nu }}a(0)\right)e^{-i \omega _{\nu }t}+i\sum _{\nu }\frac{g_{\nu }^2}{\hbar ^2 \omega _{\nu }}a(t)-i\sum _{\nu }\frac{g_{\nu }^2}{\hbar ^2 \omega _{\nu }}\int _0^te^{-i \omega _{\nu }(t- \tau )}\frac{da(\tau)}{d\tau}d\tau .
\label{eq_coll2}
\end{eqnarray}
Introducing the renormalized frequency and creation   operators, respectively,
\begin{eqnarray}
\nonumber
\Omega &=&\omega -\sum _{\nu } \frac{g_{\nu }^2}{\hbar ^2 \omega _{\nu }},\nonumber \\
\tilde{a}_{\nu}^{+}&=&a_{\nu }^{+}+\frac{g_{\nu }}{\hbar  \omega _{\nu }}a^+,
\end{eqnarray}
%Then,
%one can rewrite the Eqs. (\ref{eq_coll2}) as follows
%\begin{eqnarray}
%\frac{d}{dt}a^+(t)&=&i a^+(t)\Omega +\frac{i}{\hbar }\sum _{\nu } g_{\nu }\tilde{a}_{\nu}^{+}(0)e^{i \omega _{\nu }%t}+i\sum _{\nu }\frac{g_{\nu }^2}{\hbar ^2 \omega _{\nu }}
%\int _0^te^{i \omega _{\nu }(t- \tau )}\frac{da^{+}%%%$(\tau)}{d\tau}d\tau,\nonumber \\
%\frac{d}{dt}a(t)&=&-i a(t)\Omega -\frac{i}{\hbar }\sum _{\nu } g_{\nu }\tilde{a}_\nu(0)e^{-i \omega _{\nu }t}-i\sum %_{\nu }
%\frac{g_{\nu }^2}{\hbar ^2 \omega _{\nu }}\int _0^te^{-i \omega _{\nu }(t- \tau )}\frac{da(\tau)}{d\tau}d%\tau .
%\end{eqnarray}
we obtain from Eqs. (\ref{eq_coll2}) the Langevin type equations
\begin{eqnarray}
\frac{d}{dt}a^+(t)&=&i a^+(t)\Omega +i F^+(t)+i\int _0^tK^*(t-\tau )\frac{da^{+}(\tau)}{d\tau}d\tau, \nonumber \\
\frac{d}{dt}a(t)&=&-i a(t)\Omega -i F(t)-i\int _0^tK(t-\tau )\frac{da(\tau)}{d\tau}d\tau ,
\label{eq_Lanj}
\end{eqnarray}
with  the dissipative kernels
$$K^*(t)=\sum _{\nu } \frac{g_{\nu }^2}{\hbar ^2 \omega _{\nu }}e^{i \omega _{\nu }t},$$
$$K(t)=\sum _{\nu } \frac{g_{\nu }^2}{\hbar ^2 \omega _{\nu }}e^{-i \omega _{\nu }t},$$
 and random forces
$$F^+(t)=\sum _{\nu } F^+_{\nu }(t)=\frac{1}{\hbar }\sum _{\nu } g_{\nu }\tilde{a}_{\nu}^{+}(0)e^{i \omega _{\nu }t},$$
$$F(t)=\sum _{\nu } F_{\nu }(t)=\frac{1}{\hbar }\sum _{\nu } g_{\nu }\tilde{a}_{\nu}(0)e^{-i \omega _{\nu }t}.$$

\subsection{Fluctuation-dissipation relations}
The fluctuation-dissipation relations are the relations between
the dissipation of a collective subsystem and the fluctuations
of random forces. Those relations express the non-equilibrium
behavior of the system in terms of equilibrium or
quasi-equilibrium characteristics. They ensure that the system
approaches the  equilibrium state.

For the correlation of the random force, one can obtain
$$<<F_{\nu }^+(t)>>=<<F_{\nu }(t)>>=<<F_{\nu }^+(t)F_{\nu }^+(t)>>=<<F_{\nu }(t)F_{\nu }(t)>>=0$$
and
\begin{eqnarray}
<<F_{\nu }^+(t)F_{\nu }(\tau )>>&=&\frac{g_{\nu }^2}{\hbar ^2}<<\tilde{a}_{\nu }^+(0)\tilde{a}_{\nu }(0)>>e^{i \omega _{\nu }(t-\tau )}=\frac{g_{\nu }^2}{\hbar ^2}e^{i \omega _{\nu }(t-\tau )}n_{\nu }, \nonumber \\
<<F_{\nu }(t)F_{\nu }^+(\tau )>>&=&\frac{g_{\nu }^2}{\hbar ^2}<<\tilde{a}_{\nu }(0)\tilde{a}_{\nu }^+(0)>>e^{-i \omega _{\nu }(t-\tau )}=\frac{g_{\nu }^2}{\hbar ^2}e^{-i \omega _{\nu }(t-\tau )}[1-n_{\nu }],
\label{eq_corr}
\end{eqnarray}
where $n_{\nu }=1/\left(1+\exp\left[\frac{(\hbar  \omega _{\nu } - \mu)}{kT}\right]\right)$ is the equilibrium Fermi-Dirac
 distribution of the occupation numbers for fermions depending on the temperature $T$.
The parameters $\mu$ is the chemical potential that offers the possibility to   fix arbitrarily the initial occupation of the
 states to which the system is coupled to. In most applications shown below, we will simply assume $\mu=0$.
However, as we will see, the change of $\mu$ significantly affects the evolution.
Here, the symbol $<<...>>$ denotes the average
over the bath.
Using the expressions  (\ref{eq_corr}), one can find the fluctuation dissipation relations
\begin{eqnarray}
K^*(t-\tau )=\sum _{\nu } \frac{<<F_{\nu }^+(t)F_{\nu }(\tau )>>}{\omega _{\nu } n_{\nu }}, \\
K(t-\tau )=\sum _{\nu } \frac{<<F_{\nu }(t)F_{\nu }^+(\tau )>>}{\omega _{\nu}[1- n_{\nu }]}.
\label{eq_fluct}
\end{eqnarray}

\subsection{Analytical expressions for   occupation number}
It is convenient to introduce the spectral density $\rho (w)$ of the heat bath excitations, which allows us to
replace the sum over different two-level systems $\nu $ by integral over the frequency:
$\sum_\nu...\to \int_{0}^{\infty}dw\rho (w)...$. Let us consider the following spectral function \cite{M1}
\begin{eqnarray}
\dfrac{g_{\nu }^2}{\hbar ^2 w_{\nu }}\to\dfrac{\rho (w) g^2(w)}{\hbar ^2 w}=\dfrac{1}{\pi }g_{0}\dfrac{\gamma ^2}{\gamma^2+w^2},
\label{eq_222}
\end{eqnarray}
where the memory time $\gamma ^{-1}$ of the dissipation is inverse to the bandwidth of the heat bath
excitations which are coupled to collective subsystem. This is the Ohmic dissipation with the Lorenzian
cutoff (Drude dissipation). The relaxation time of the heat bath should be much less than   the
characteristic collective time, i.e.  $\gamma\gg \omega$.

Employing Eq. (\ref{eq_222}), we obtain the  expressions for the dissipative kernels:
\begin{eqnarray}
K^*(t)=\frac{1}{\pi}g_{0}\gamma^2\int_{0}^{\infty}dw \frac{e^{i w t}}{\gamma^2+w^2}=
\frac{1}{2}  g_{0} \gamma  e^{-\gamma t}+\frac{i}{\pi}g_{0}\gamma ^2
\int_{0}^{\infty}dw \frac{\sin[w t]}{\gamma^2+w^2}, \nonumber \\
K(t)=\frac{1}{\pi}g_{0}\gamma^2\int_{0}^{\infty }dw \frac{e^{-i w t}}{\gamma^2+w^2}=
\frac{1}{2}  g_{0} \gamma   e^{-\gamma t}-\frac{i}{\pi}g_{0}\gamma ^2
\int_{0}^{\infty}dw \frac{\sin[w t]}{\gamma^2+w^2}.
\label{eq_2222pm}
\end{eqnarray}
To  leading order in $g_0$,
these dissipative kernels are approximated in the following way \cite{M1}
\begin{eqnarray}
K^*(t)=K(t)=\frac{2}{\pi}g_{0}\gamma^2\int_{0}^{\infty}dw \frac{\cos(wt)}{\gamma^2+w^2}=
  g_{0} \gamma  e^{-\gamma t}.
%\int_{0}^{\infty}dw \frac{\sin[w t]}{\gamma^2+w^2}, \nonumber \\
%K(t)=\frac{1}{\pi}g_{0}\gamma^2\int_{0}^{\infty }dw \frac{e^{-i w t}}{\gamma^2+w^2}=
%\frac{1}{2}  g_{0} \gamma   e^{-\gamma t}-\frac{i}{\pi}g_{0}\gamma ^2
%\int_{0}^{\infty}dw \frac{\sin[w t]}{\gamma^2+w^2}.
\label{eq_2222pm}
\end{eqnarray}
Besides the specific nature of the bath (see Fig. \ref{figure0}), setting to zero the imaginary part of $K(t)$
is the only approximation that is made in the present work. Comparing with other approach \cite{Lac14}, it could
be shown that this approximation does   not influence the non-equilibrium evolution in the weak coupling regime.
In the limit $\gamma\to\infty$, the dissipative kernels  (\ref{eq_2222pm})  have a well-known Markovian form
\begin{eqnarray}
K(t)=2 g_0 \delta(t).
\label{eq_2222pmg}
\end{eqnarray}
For the sake of simplicity, in further analytical calculations  we use  the  dissipative kernels (\ref{eq_2222pm}).

For the average occupation number $n(t)$, one can obtain (see Appendix A):
\begin{eqnarray}
n(t)=<<a^+(t){\ }a(t)>> =\tilde{A}^*(t)\tilde{A}(t) <<a^{+}(0){\ }a(0)>> +<<\tilde{F}^+(t)\tilde{F}(t)>>
\label{eq_nt}
\end{eqnarray}
with the explicit expression
\begin{eqnarray}
<<\tilde{F}^+(t)\tilde{F}(t)>>=
\frac{1}{\pi }g_0\gamma ^2\int^{\infty}_{0} dw\frac{w}{\gamma ^2+w^2}
B_w^*(t)B_w(t)n_\nu(\omega),
\end{eqnarray}
where
\begin{eqnarray}
B_w^*(t)=\frac{e^{i t w} (z_1-z_2) (i w+\gamma )+e^{t z_1} (-i w+z_2) (z_1+\gamma )+i e^{t z_2} (w+i z_1) (z_2+\gamma )}{(w+i z_1) (z_1-z_2) (w+i z_2)},
\end{eqnarray}
and
\begin{eqnarray}
\tilde{A}^*(t)=e^{t z_1}\frac{z_1+ \gamma -i g_0 \gamma }{z_1-z_2}+e^{t z_2}\frac{z_2+ \gamma -i g_0 \gamma }{z_2-z_1}.
\end{eqnarray}
The expressions for $z_1$ and $z_2$ are given in Appendix A.
The equilibrium population of the fermionic oscillator is obtained by taking   $t\to\infty$ limit in (\ref{eq_nt}).
The first term of $B_w^{*}(t)$  contributes only  to  the asymptotic value of the occupation number:
\begin{eqnarray}
n(t\to\infty)=<<\tilde{F}^+(t)\tilde{F}(t)>>|_{t\to\infty}=\frac{1}{\pi }g_0\gamma ^2\int^{\infty}_{0} dw  \frac{w}{|w+i z_1|^2|w+i z_2|^2}n_\nu(\omega).
\label{eq_ntasymp}
\end{eqnarray}

If  at $t=0$ the collective subsystem is in the ground state,
then  $\ll a^{+}(0){\ }a(0)\gg=0$ and
%\begin{eqnarray}
%n(t)=\frac{1}{\pi }g_0\gamma ^2\int^{\infty}_{0} dw\frac{1}{\gamma ^2+w^2}\frac{w}{1+\exp\left[\frac{\hbar
%w}{kT}\right]}B_w^*(t)B_w(t).
%\label{nTime}
%\end{eqnarray}
\begin{eqnarray}
n(t)=\frac{1}{\pi }g_0\gamma ^2\int^{\infty}_{0}dw \frac{w}{\gamma ^2+w^2}B_w^*(t)B_w(t)n_\nu(\omega).
\label{nTime}
\end{eqnarray}
This compact expression is rather interesting to identify the effect of the Fermi statistics.
Indeed, three contributions to the evolution
can be clearly separated. First, we easily identify the influence of the specific spectral function.
Here, we will assume large $\gamma$ value to insure
that the system is coupled to a large set of states in the bath.
The second contribution stems from $B_w^*(t)B_w(t)$.
%and contains the specific treatment of the bath considered here.
This term only depends on the original equation of motion
[see Eqs. (\ref{eq_coll}) and (\ref{eq_intr})].
These two factors  contains the specific treatment of the bath considered here.
Due to the two-level nature of the bath, their contributions turn  out to be identical to those of bosonic bath.
The only direct effect of the fermionic nature of the bath  is  the Fermi-Dirac statistics that determines
the initial occupancies $n_\nu(\omega)$.

Note that we will also compare the results with the reference case of a system coupled to a bosonic bath. This case is simply
obtained by employing   the following replacement
$$\frac{1}{1+\exp\left[\frac{\hbar w}{kT}\right]}\rightarrow \frac{1}{\exp\left[\frac{\hbar w}{kT}\right]-1},$$
in Eq. (\ref{eq_nt}). One can obtain the expression for the occupation number of  bosonic  system
(collective bosonic oscillator plus   internal  bosonic   oscillators).
This   expression  for bosonic system  was firstly derived  in  Ref. \cite{M1}.

\subsection{Analytical expressions for  level population at $g_0\ll1$ and at high and low temperatures}
\label{sec:asympt}

Retaining terms to leading order in $g_0$, we obtain at $g_0\ll 1$ that
\begin{eqnarray}
z_1=-\gamma +i g_0 \gamma ,\\\nonumber
z_2=i \Omega -g_0 \gamma ,
\end{eqnarray}
$$\tilde{A}^*(t)= e^{z_1t} \frac{(z_1+\gamma -i \gamma  g_0)}{z_1-z_2}+e^{z_2t} \frac{(z_2+\gamma -i \gamma  g_0) }{z_2-z_1} =e^{z_2t},$$
$$\tilde{A}^*(t)\tilde{A}(t)=e^{(z_2+z_2^*)t}=e^{-2 g_0  \Omega t},$$
and
\begin{eqnarray}
n(t)&=&\ll a^{+}(0)a(0)\gg e^{-2 g_0  \Omega t}+\frac{1}{\pi }g_0\gamma ^2 \int^{\infty}_{0}dw\frac{1}{\gamma ^2+w^2}\frac{w}{1+\exp\left[\frac{\hbar  w}{ kT}\right]}\times \\\nonumber
&\times &\left\{ f_1+ f_2e^{-2 \gamma t}+f_3e^{-2 g_0 \Omega t}+2{\rm Re}[ f_4e^{(z_1-i w)t}]+2{\rm Re}[f_5e^{(z_2-i w)t}]
+2{\rm Re}[f_6e^{(z_1+z_2^{*})t}]] \right\}.
\end{eqnarray}
The analytical expressions for the  functions  $f_i$  are presented in Appendix B.
%\subsection{High- and low-temperature  limits}
Another good presentation is as follows
\begin{eqnarray}
n(t)&=&\ll a^{+}(0)a(0)\gg e^{-2 g_0 \Omega t}+n(t\to \infty)[1+e^{-2 g_0 \Omega t}],
\end{eqnarray}
\begin{eqnarray}
n(t\to\infty)=\frac{1}{\pi }g_0 \gamma ^2 \int_{0}^{\infty}
dw\frac{w}{1+\exp\left[\frac{\hbar  w}{kT}\right]}\frac{1}{\left[ w^2- g_0 w \gamma +\left(1+ g_0^2\right) \gamma ^2\right] \left[  w^2-2 w \Omega +\left(1+g_0^2\right) \Omega ^2\right]}.
\label{eq_asym1}
\end{eqnarray}
At high   temperatures ($kT\gg \hbar\omega$) and $\hbar\gamma\gg kT$,
 the main contribution to the integral (\ref{eq_asym1}) arises  in the neighborhood of $|w|\approx\Omega$ and
%$$\dfrac{1}{1+\exp(\hbar w/T)}\approx\dfrac{1}{2}-\dfrac{\hbar w}{4T}$$
 yields the usual Fermi-Dirac population
\begin{eqnarray}
n_F=\frac{1}{\exp\left[\frac{\hbar \Omega }{kT}\right]+1},
\label{eq_fermi}
\end{eqnarray}
showing that the bath imposes its temperature to the system. Similarly, for bosonic
systems, in this limit, we obtain
\begin{eqnarray}
n_B=\frac{1}{\exp\left[\frac{\hbar \Omega }{kT}\right] -1}.
\label{eq_boson}
\end{eqnarray}

 At $\gamma\to \infty$ and $g_0\ll 1$,
 \begin{eqnarray}
n(t\to\infty)=\frac{g_0}{\pi }  \int_{0}^{\infty}
dw\frac{w}{\exp\left[\frac{\hbar  w}{kT}\right]\pm 1}\frac{1}{ w^2-2 w \Omega + \Omega ^2}.
\label{eq_g00}
\end{eqnarray}

 At the low-temperature limit ($kT\ll \hbar\omega$), in (\ref{eq_g00}) one can
employ   the following expansion:
\begin{eqnarray}
\frac{1}{\left[  w^2-2 w \Omega + \Omega ^2\right]}
\approx
\frac{1}{\Omega^2}\left[1+2\frac{w}{\Omega}+3\left(\frac{w}{\Omega}\right)^2\right].
\label{eq_approx2}
\end{eqnarray}
Then, we obtain
\begin{eqnarray}
n(t\to\infty)=\frac{g_0}{\pi }\frac{1}{\Omega ^2} \int_{0}^{\infty}
dw\frac{w}{\exp\left[\frac{\hbar  w}{kT}\right]\pm 1}\left[1+2\frac{w}{\Omega}+3\left(\frac{w}{\Omega}\right)^2\right].
\label{eq_asym}
\end{eqnarray}
from which
one can derive  the following analytical expressions
 \begin{eqnarray}
n_B(t\to\infty)=\frac{g_0}{\pi}\left(\frac{kT}{\hbar\Omega}\right)^2 \left[\zeta(2)+4\frac{kT}{\hbar\Omega}\zeta(3)+18\left(\frac{kT}{\hbar\Omega}\right)^2\zeta(4)\right]
\label{nBT0}
\end{eqnarray}
and
\begin{eqnarray}
n_F(t\to\infty)=\frac{g_0}{\pi}\left(\frac{kT}{\hbar\Omega}\right)^2 \left[\frac{1}{2}\zeta(2)+3\frac{kT}{\hbar\Omega}\zeta(3)+\frac{63}{4}\left(\frac{kT}{\hbar\Omega}\right)^2\zeta(4)\right],
\label{nFT0}
\end{eqnarray}
for bosonic and fermionic systems, respectively.
Here, $\zeta(n)$ is the Reman zeta-function.
We see
that the $n_{F,B}(t\to\infty)$  is nearly linear in $g_0$ in this regime.
At $T\to 0$,
 \begin{eqnarray}
n_F(t\to\infty)/n_B(t\to\infty)\to \frac{1}{2}.
\label{eq_asym8}
\end{eqnarray}
This limit will be explicitly studied in Sec. III.B.

\subsection{Friction and diffusion coefficients}
In order to calculate the friction and diffusion coefficients, we employ Eq. (\ref{eq_aaa}) and
obtain the following equations:
\begin{eqnarray}
\frac{d}{dt}<<a^+(t)>>&=&\frac{d\tilde{A}^*(t)/dt}{\tilde{A}^*(t)}<<a^+(t)>>, \nonumber\\
\frac{d}{dt}<<a(t)>>&=&\frac{d\tilde{A}(t)/dt}{\tilde{A}(t)}<<a(t)>>, \nonumber\\
\frac{d}{dt}<<a^+(t)a(t)>>&=&\frac{d[\tilde{A}^*(t)\tilde{A}(t)]/dt}{\tilde{A}^*(t)\tilde{A}(t)}<<a^+(t)a(t)>>
-\frac{d[\tilde{A}^*(t)\tilde{A}(t)]/dt}{\tilde{A}^*(t)\tilde{A}(t)}<<\tilde{F}^+(t)\tilde{F}(t)>> \nonumber\\
&+&\frac{d<<\tilde{F}^+(t)\tilde{F}(t)>>/dt}{\tilde{A}^*(t)\tilde{A}(t)}, \nonumber\\
\frac{d}{dt}<<a(t)a^+(t)>>&=&\frac{d[\tilde{A}(t)\tilde{A}^*(t)]/dt}{\tilde{A}(t)\tilde{A}^*(t)}<<a(t)a^+(t)>>-
\frac{d[\tilde{A}(t)\tilde{A}^*(t)]/dt}{\tilde{A}(t)\tilde{A}^*(t)}<<\tilde{F}(t)\tilde{F}^+(t)>> \nonumber\\
&+&\frac{d<<\tilde{F}(t)\tilde{F}^+(t)>>/dt}{\tilde{A}(t)\tilde{A}^*(t)}, \nonumber\\
\frac{d}{dt}<<a^+(t)a^+(t)>>&=&\frac{d[\tilde{A}^*(t)\tilde{A}^*(t)]/dt}{\tilde{A}^*(t)\tilde{A}^*(t)}<<a^+(t)a^+(t)>>, \nonumber\\
\frac{d}{dt}<<a(t)a(t)>>&=&\frac{d[\tilde{A}(t)\tilde{A}(t)]/dt}{\tilde{A}(t)\tilde{A}(t)}<<a(t)a(t)>>.
\end{eqnarray}
As seen from the structure of these equations,  the dynamics
of collective subsystem is determined by the time-dependent friction
\begin{eqnarray}
\lambda(t)=-\frac{1}{2}\frac{d\ln[\tilde{A}^*(t)\tilde{A}(t)]}{dt}=-\frac{1}{2}\frac{d[\tilde{A}^*(t)\tilde{A}(t)]/dt}{\tilde{A}^*(t)\tilde{A}(t)}
=-\frac{1}{2}[\frac{d\tilde{A}^*(t)/dt}{\tilde{A}^*(t)}+\frac{d\tilde{A}(t)/dt}{\tilde{A}(t)}]
\end{eqnarray}
\label{eq_fric}
and diffusion
\begin{eqnarray}
D_{a^+a}(t)&=&-\frac{1}{2}\frac{d[\tilde{A}^*(t)\tilde{A}(t)]/dt}{\tilde{A}^*(t)\tilde{A}(t)}<<\tilde{F}^+(t)\tilde{F}(t)>>+\frac{1}{2}\frac{d<<\tilde{F}^+(t)\tilde{F}(t)>>/dt}{\tilde{A}^*(t)\tilde{A}(t)}\\ \nonumber
&=&\lambda(t)<<\tilde{F}^+(t)\tilde{F}(t)>>+\frac{1}{2}\frac{d<<\tilde{F}^+(t)\tilde{F}(t)>>/dt}{\tilde{A}^*(t)\tilde{A}(t)},
\label{eq_diff}
\end{eqnarray}
\begin{eqnarray}
D_{aa^+}(t)&=&-\frac{1}{2}\frac{d[\tilde{A}^*(t)\tilde{A}(t)]/dt}{\tilde{A}^*(t)\tilde{A}(t)}<<\tilde{F}(t)\tilde{F}^+(t)>>+\frac{1}{2}\frac{d<<\tilde{F}(t)\tilde{F}^+(t)>>/dt}{\tilde{A}^*(t)\tilde{A}(t)}=\\ \nonumber
&=&\lambda(t)<<\tilde{F}(t)\tilde{F}^+(t)>>+\frac{1}{2}\frac{d<<\tilde{F}(t)\tilde{F}^+(t)>>/dt}{\tilde{A}^*(t)\tilde{A}(t)}
\label{eq_diff2}
\end{eqnarray}
coefficients.
Therefore, we have obtained the local in time Markovian-type  equations for the first and second moments, but with a
general form of transport coefficients which explicitly depend
on time.
The non-Markovian effects are taken into consideration through this time dependence.
It can be shown that the appropriate
%grand-canonical
equilibrium distribution is achieved in the course of
 time evolution. At $t\to\infty $ the system reaches the equilibrium
state ($\frac{d}{dt}<<a^+>>=\frac{d}{dt}<<a>>=\frac{d}{dt}<<a^+a>>=\frac{d}{dt}<<aa^+>>=\frac{d}{dt}<<a^+a^+>>
=\frac{d}{dt}<<aa>>=0$) and the asymptotic diffusion
coefficients can be derived from the above expressions:
\begin{eqnarray}
D_{a^+a}(t\to\infty)=\lambda(t\to\infty)n(t\to\infty),
\label{eq_diffasym}
\end{eqnarray}
\begin{eqnarray}
D_{aa^+}(t\to\infty)=\lambda(t\to\infty)[1-n(t\to\infty)],
\label{eq_diff2asym}
\end{eqnarray}
where the asymptotic friction coefficient
\begin{eqnarray}
\lambda(t\to\infty)=-\frac{1}{2}[z_2+z_2^*].
\label{eq_fricasym}
\end{eqnarray}
It is interesting to mention that the time evolution of the friction coefficient and consequently its asymptotic
limit do not depend  on the fact that the bath is composed of fermions or bosons. However, the specific quantum natures
of the bath enter into the diffusion coefficients through the appearance of occupation probabilities in
Eqs. (\ref{eq_diffasym}) and (\ref{eq_diff2asym}).

The asymptotic diffusion and friction coefficients
are related by the well-known fluctuation-dissipation
relations connecting diffusion and damping constants.
Note that
$$\frac{d[\tilde{A}^*(t)\tilde{A}^*(t)]/dt}{\tilde{A}^*(t)\tilde{A}^*(t)}|_{t\to\infty}=2\frac{d\tilde{A}^*(t)/dt}{\tilde{A}^*(t)}|_{t\to\infty}= 2z_2,$$
$$\frac{d[\tilde{A}(t)\tilde{A}(t)]/dt}{\tilde{A}(t)\tilde{A}(t)}|_{t\to\infty}=2\frac{d\tilde{A}(t)/dt}{\tilde{A}(t)}|_{t\to\infty}=2z_2^*,$$
and at $g_0\ll 1$
$$\lambda(t\to\infty)=g_0\Omega.$$

\section{\protect\bigskip Calculated results}
The population, diffusion and friction coefficients depend on the parameters $\omega$, $g_0$, and $\gamma$.
The value of $\gamma$ should be taken to hold the condition $\gamma\gg\omega$  or $\gamma\gg\Omega$ .
We set $\gamma/\Omega=$ 12 while the renormalized frequency is taken $\hbar\Omega=$1 MeV.
To calculate the level population, friction and diffusion coefficients,
 we use formulas (\ref{eq_nt}), (\ref{eq_ntasymp}), (\ref{eq_fricasym}), and (\ref{eq_diffasym}).

\begin{figure}[h]
\includegraphics[scale=1.1]{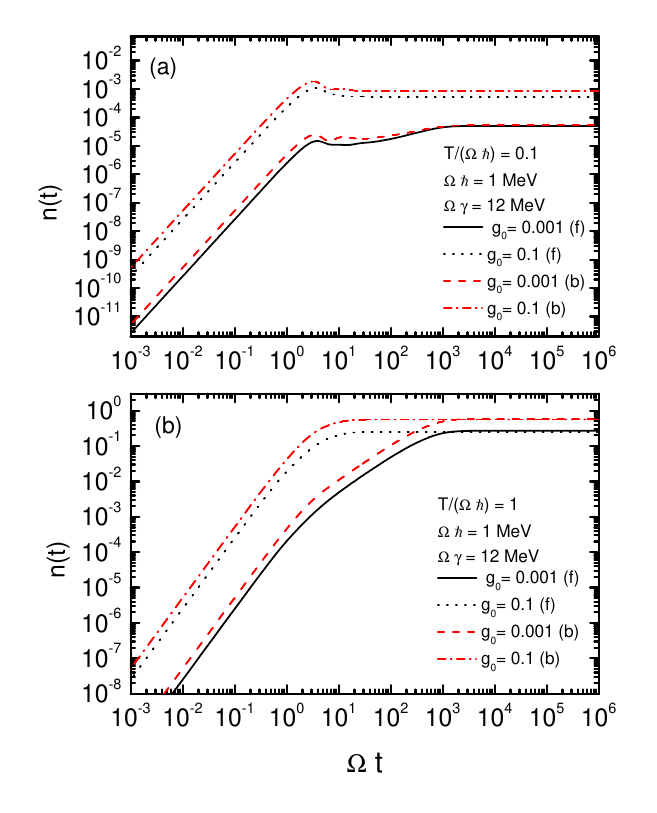} \hspace*{0.5cm}
\includegraphics[scale=1.1]{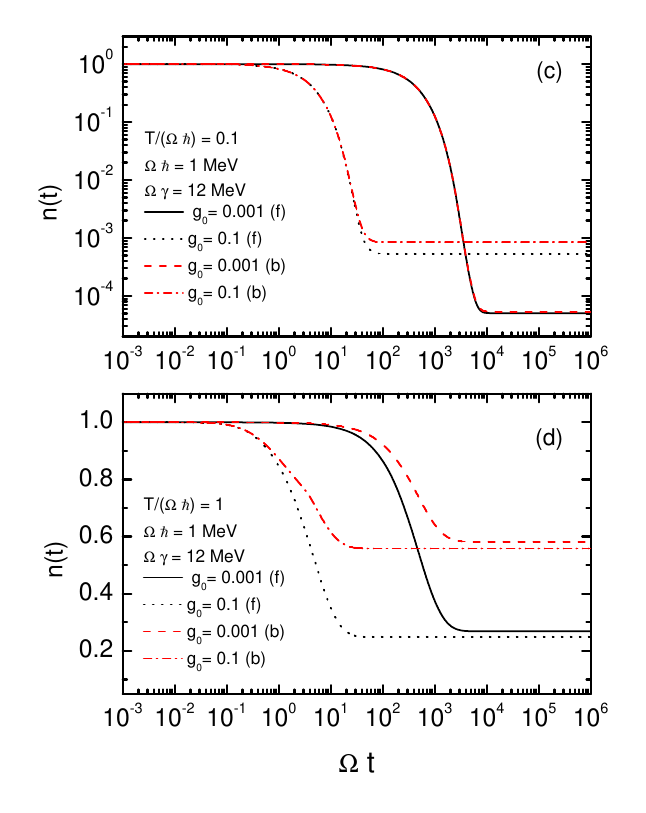}
\caption{(Color online) The calculated dependencies of the  average  occupation numbers $n(t)$ on time $t$ for the fermionic and
bosonic systems, respectively, labeled by "(f)" and "(b)". The results for different coupling constants $g_0$ and temperatures $T$ are indicated in the plots.
Results (a) and (b) [(c) and (d)] correspond to an initially unoccupied, i.e. $n(t=0)$=0 [occupied, i.e. $n(t=0)$=1] system state.}
\label{n_time}
\end{figure}

\subsection{Non-equilibrium properties}

For bosonic and fermionic systems, the time dependence of occupation numbers $n(t)$   are shown in Fig. \ref{n_time}.
Note that here a zero chemical potential is assumed.
In Fig.  \ref{n_time}, the collective subsystem is initially unoccupied with $n(0)=$0 (left hand side), or occupied with $n(0)=$1 (right hand side).
If the state is initially unoccupied and at low temperature,
we see that, before reaching an equilibrium, the non-equilibrium evolution is affected by the fermionic nature of the bath.
More precise analyze shows that some oscillations of the occupation numbers exist  in the bosonic case.
In the fermionic case, these oscillations are smaller.
When temperature increases the difference between the Fermi and Bose systems seems to be
washed out by the thermal fluctuations.
After some transient time, which depends on the coupling strength, the temperature, spectral function, and
the level populations reach their equilibrium values.
For the bosonic and fermionic systems,  we observe that the transient times are almost the same  and independent
of the initial collective  subsystem occupancy.
In particular, we see in Fig. \ref{n_time} that the transient time increases with decreasing  coupling strength $g_0$.
Thus, the transition occurs  faster in the case of larger coupling strength.

\subsection{Asymptotic properties}
\label{sec:lowglowT}

\begin{figure}[h]
\includegraphics[scale=1.]{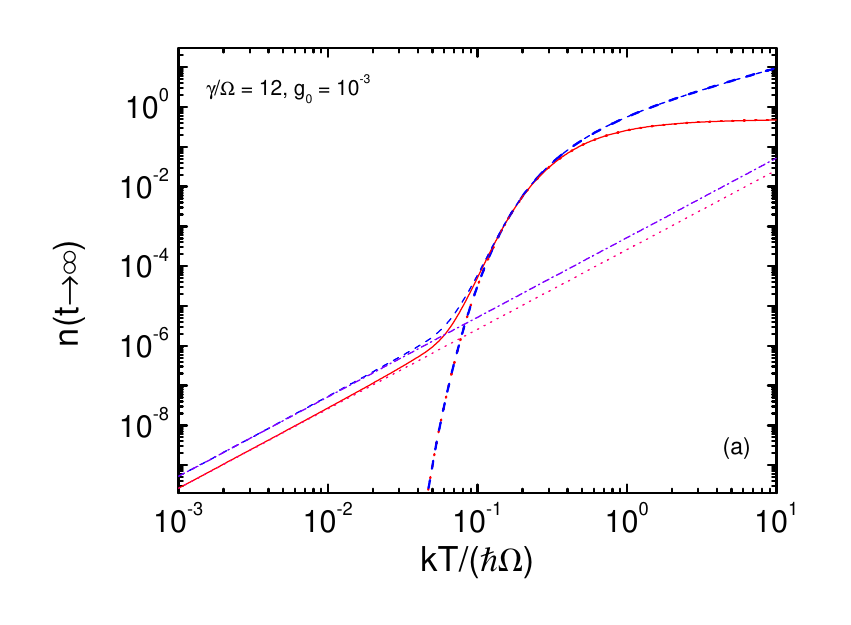}
\includegraphics[scale=1.]{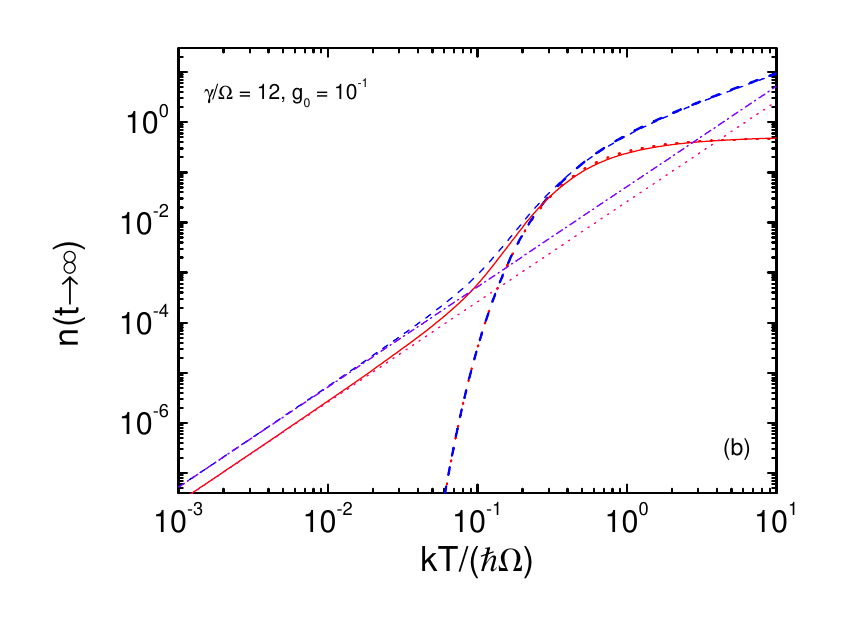}
\caption{(Color online) The calculated dependencies of the asymptotic average
occupation numbers $n(t\to\infty)$   for fermionic system (solid lines) and
for bosonic  system (dashed lines) on temperature $T$.
The $g_0=0.001$ (a) and $g_0=0.1$ (b) cases are presented.
The occupation numbers for the Fermi-Dirac and Bose-Einstein distributions are presented by thick dotted lines
and thick dashed  lines, respectively.
The leading order expressions obtained for fermions and bosons in power of $(k T)/(\hbar \Omega)$,
Eqs. (\ref{nBT0}) and (\ref{nFT0}), are presented by dotted and dash-dotted lines, respectively.
}
\label{n_aver}
\end{figure}

We see in Fig. \ref{n_time} that after some relaxation time,
the collective subsystem reaches an equilibrium.
Before equilibrium, at low temperature [$kT/(\hbar\Omega)=0.1$], the   occupation numbers increase
with increasing $g_0$ for both fermionic and bosonic systems. This means that at
low temperature the friction strongly influences the dynamics of occupation numbers.
At intermediate temperature [$kT/(\hbar\Omega)=1$],  the ratio between the asymptotic
occupation numbers for bosonic and fermionic  systems in this figure is about of factor  $2$   for   different values of $g_0$.
As   discussed in Sec. II.D, it is anticipated that at high temperature and weak coupling   the asymptotic
equilibrium essentially reflects the fermionic or bosonic nature of the bath that is imposed to the collective subsystem.
Deeper insight in the equilibrium properties can be obtained using Eq. (\ref{eq_asym}) for the equilibrium occupation number.
For two coupling strengths  $g_0=0.1$ and $g_0=0.001$,
the dependencies of   $n(t\to\infty)$  on temperature $T$  are systematically investigated in Fig. \ref{n_aver}.
The Fermi-Dirac and Bose-Einstein distributions are also shown as the references.
At high temperature, the occupation numbers  for the fermionic and bosonic systems correspond to the
 Fermi-Dirac and Bose-Einstein populations when the temperature
is above a certain threshold and the equilibrium properties of the system is imposed by the heat-bath.
This is an indirect argument of the correctness of our method.
At low temperatures, there are noticeable deviations from the
usual Fermi-Dirac and Bose-Einstein   populations behavior.
For the bosonic system, this effect was firstly found in Ref. \cite{M1}.

\begin{figure}[h]
\begin{center}
\includegraphics[scale=1.]{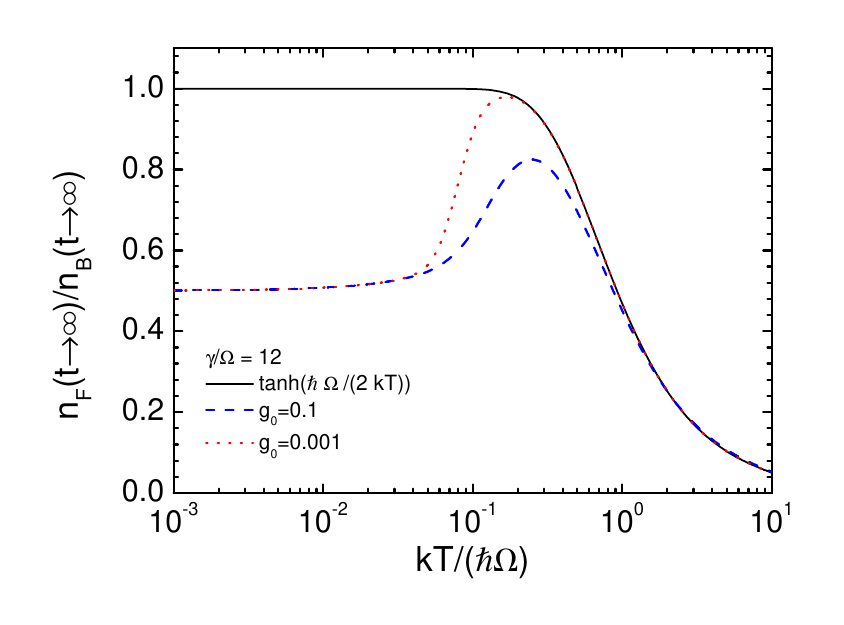}
\end{center}
\caption{(Color online) Ratio $n_F(t\to\infty) / n_B(t\to\infty)$ displayed as  function of temperature for $g_0=0.1$ (dashed  line) and
$g_0= 0.001$(dotted line). The curve ${\rm tanh} [\hbar \Omega /(2 k T)]$ (solid line) is also shown for the reference.
}
\label{fig:ratio}
\end{figure}

In Fig. \ref{fig:ratio}, the calculated ratio $n_F(t\to\infty) / n_B(t\to\infty)$
is compared to the completely thermalized limit as a function of temperature.
At  [$kT/(\hbar\Omega)=1$], both fermionic and bosonic systems are above the threshold mentioned and  the factor $1/2$
can be simply  explained using the approximate formula:
\begin{eqnarray}
\frac{n_F(t\to\infty)}{n_B(t\to\infty)} & \simeq & {\rm tanh} \left(\frac{\hbar \Omega}{2 k T} \right)
\end{eqnarray}
leading to a value of $0.46$.
 We see that above of certain  temperature,
which depends on the coupling strength,
the occupation numbers follow  the statistics imposed
by the bath for all coupling strengths considered.
%the bath imposes its statistics, fermionic or bosonic,
%to the collective subsystem leading
%to a noticeable deviation from the  anticipated behavior (\ref{eq_fermi}).
The low temperature regime is less straightforward to understand. In this case, the quantum fluctuations compete with the statistical
fluctuations and a clear deviation from the thermal equilibrium prescription is observed.  In particular, we see that the transition
temperature between quantal and statistical effects strongly depends on the interaction strength.
%
%Expressions (\ref{nBT0}) and (\ref{nFT0}) provide us the interesting information on the low temperature limit. First,
%We see in Fig.3
%that the $n_{F,B}(t\to\infty)$  is nearly linear in $g_0$ in the weak coupling regime.
%Such effect is indeed observed in by comparing the
%two cases presented in left and right panel of Fig. \ref{n_aver}.
The low temperature regime is
significantly affected by the nature of the bath, leading to a factor of $1/2$ between the  $n_F(t\to\infty)$ and $n_B(t\to\infty)$ at
$T\to 0$.
This aspect has been analytically studied with Eqs. (\ref{nBT0}) and (\ref{nFT0}). The
result of these equations
are shown in Fig. \ref{n_aver} and perfectly match the numerical result in the low
temperature regime (below $kT/\hbar\Omega=0.1$). The influence of quantum statistics is further
illustrated in the ratio shown in Fig. 4.
As   seen, this will have a direct influence   the diffusion process.
Note that
%the results are not sensitive to the  $\gamma$.
because the calculations are not sensitive to the de-excitation constant $\gamma$,
there is no influence of the spectral function.

\begin{figure}[htbp]
\begin{center}
\includegraphics[scale=1.0]{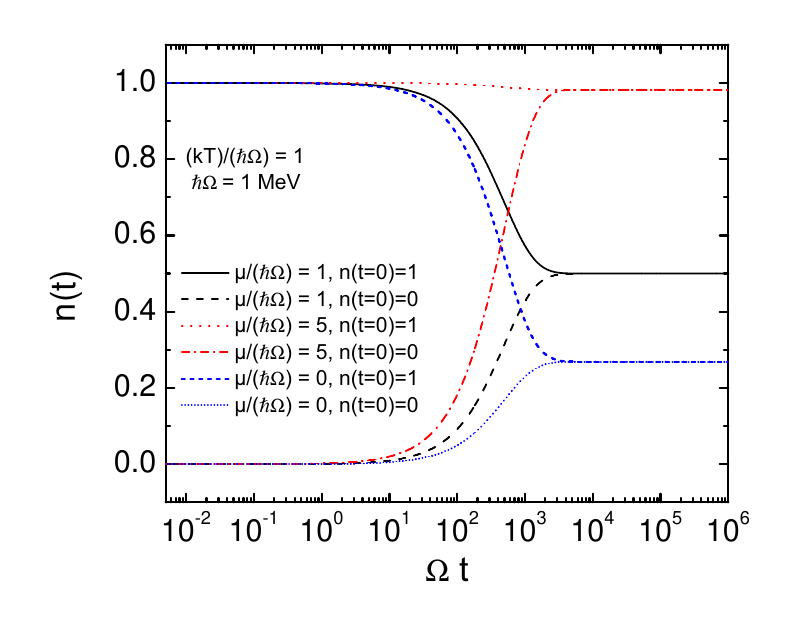}
\end{center}
%\vspace{-2.3cm}
\caption{(Color online)
The calculated dependence of the  average  occupation number $n(t)$ on time $t$ for the fermionic system.
The results for different chemical potentials $\mu$ and initial average  occupation number $n(0)$ are indicated.}
\label{n_mu}
\end{figure}

\subsection{Non-zero chemical potential and Pauli principle effect}

At given temperature, it is expected that a change of the chemical potential
significantly affects the asymptotic behavior and  the overall dynamics in the fermionic case.
When the chemical potential becomes comparable to or bigger than the state energy $\hbar \Omega$
and the collective subsystem is already occupied at initial time, we anticipate that it will not decay towards the bath due to the fact that the bath states are already
occupied at the relevant energy. This situation is depicted in   Fig. \ref{figure0}. Examples of time-evolution with different initial
chemical potentials for the bath are shown in Fig. \ref{n_mu} for the collective subsystem that is either fully occupied or fully empty. Since we are above of
the threshold, we observe that the asymptotic behavior is independent of the initial conditions for the collective subsystem. The Pauli principle effect is perfectly seen
in the case of $\mu/(\hbar \Omega) =5$ and a fully occupied collective subsystem at initial time. The collective subsystem occupation remains almost constant and
equal to its initial value.

\subsection{Transport properties: fluctuation and dissipation}
%\begin{figure}[h]
%\includegraphics[scale=0.9]{figure6.pdf}
%\caption{The calculated dependence of the asymptotic friction coefficient on the coupling constant $g_0$.}
%\label{n_asympt1}
%\end{figure}

\begin{figure}[h]
\includegraphics[scale=1.0]{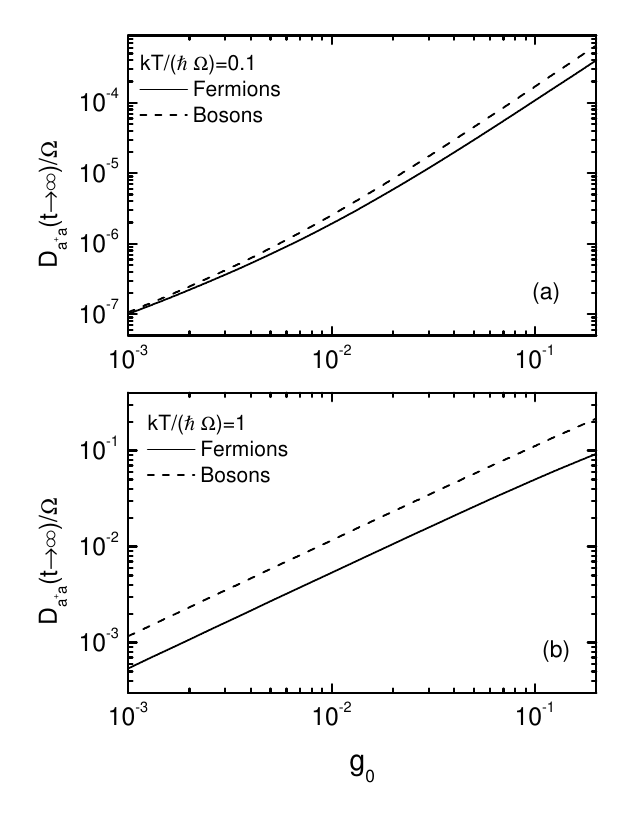}
\caption{The calculated dependence of the asymptotic diffusion coefficient on the coupling constant $g_0$.
The results for the fermionic and bosonic systems at different temperatures   are indicated.}
\label{n_asympt2}
\end{figure}
The asymptotic
%friction and
diffusion coefficients as  the functions of $g_0$ are shown in Fig.
%\ref{n_asympt1} and
\ref{n_asympt2}.
As already mentioned, the friction is the same for the Fermi or  Bose systems. Therefore, we do not expect
that  the dissipative aspects strongly depend on the nature of the particles.
An indirect proof of this was the fact that the time-scale
before reaching an equilibrium was weakly dependent  on the   statistical properties.
Both   friction and diffusion exhibit a monotonic increase with increasing coupling strength.

At low temperature, the difference between the asymptotic  diffusion coefficients for
fermionic   and  bosonic  systems is small but increases slightly with increasing $g_0$  [upper part of Fig. \ref{n_asympt2}].
At high temperature, the deviation between fermionic   and bosonic systems is
larger and almost the same for all coupling constants [lower part of Fig. \ref{n_asympt2}].
The difference between   two cases considered here directly stems from the bounded values of the
occupation numbers in the Fermi systems. Because $0 \le n(t=\infty) \le 1$, from Eqs. (\ref{eq_diffasym}) and (\ref{eq_diff2asym})
we see that the  diffusion coefficients will verify
\begin{eqnarray}
D_{a^+a}(t\to\infty) \le \lambda(t\to\infty), ~~~D_{aa^+}(t\to\infty) \le \lambda(t\to\infty).
\end{eqnarray}
 In bosonic systems, similar equations can be derived except that $D_{aa^+}(t\to\infty)=\lambda(t\to\infty)[1+n(t\to\infty)]$, and
  there is no restriction on the value of $n(t\to\infty)$. This quenching of the diffusion coefficient
 reflects the difference between the Fermi and Bose heat baths. At any temperature, the reduction factor can
 be directly estimated as
\begin{eqnarray}
\frac{D^{\rm Fermi}_{a^+a}(t\to\infty) }{D^{\rm Bose}_{a^+a}(t\to\infty)} = \frac{n_F (t\to\infty)}{n_B (t\to\infty)}
\end{eqnarray}
that is seen in Fig. \ref{fig:ratio}. As  shown in this figure, the fluctuations are reduced in the fermionic case
compared to the bosonic case both at high and low temperature. This leads to a rather subtle effect of the nature of the bath.
At high temperature, the quenching directly reflects the thermal fluctuations that are smaller in the Bose systems compared
to the Fermi systems. At lower temperature, the main effect  is a difference in the quantal zero-point motions.
% that induce  different low energy limits.

\section{\protect\bigskip Summary}
The non-Markovian quantum  Langevin equations and fluctuation-dissipation
relations are derived for the fermionic system (the collective fermionic oscillator plus   fermionic heat-bath).
The explicit expressions for the time-dependent level population, friction
and diffusion coefficients are derived  for the case of
linear coupling  between the collective and internal  subsystems.
%We have presented a detailed analysis of the connections between fermionic and
%bosonic Hamiltonians.
The results of  numerical calculations of diffusion and friction coefficients and
level population  are also shown and compared to the bosonic case. At high
temperature, regardless of the strength of  coupling between the bath and collective  subsystem,
the collective  subsystem relaxes towards occupation numbers imposed by the bath nature (bosonic or femionic).
At sufficiently low temperatures, the effect
of the coupling is macroscopically observable as the deviation of level population
from the Fermi-Dirac  population. In this case the quantum fluctuations plays a significant role.
In particular, a quenching of   the asymptotic population compared to bosons systems
is anticipated. This  affects the diffusion process while leaving the
friction unchanged.    With the recent progress in manipulating Fermi gas, it might be interesting to
explore this effect experimentally.

\section*{Acknowledgment}
This work was supported by RFBR. The IN2P3(France)-JINR(Dubna) Cooperation
Program is gratefully acknowledged.

\section*{Appendix A}

To find the solution of  Eq.~(\ref{eq_Lanj}),   we apply the Laplace transform:
\begin{eqnarray}
A^{+}(z)=a^{+}(0)\frac{1-i K(z)}{z -i \Omega  -z i K(z)}+F^{+}(z)\frac{i}{z -i \Omega  -z i K(z)},
\label{eq_apl}
\end{eqnarray}
where $A^{+}(z)=L\{a^{+}(t)\}$  and $F^{+}(z)=L\{F^{+}(t)\}$.

Substituting the expression (14) for the kernel  into  (\ref{eq_apl})
and employing the roots
$$z_1=\frac{1}{2} \left(-\gamma +i \Omega +i \gamma g_0-\sqrt{4 i \gamma  \Omega +(\gamma -i \Omega -i \gamma  g_0)^2}\right)$$
and
$$z_2=\frac{1}{2} \left(-\gamma +i \Omega +i \gamma  g_0+\sqrt{4 i \gamma  \Omega +(\gamma -i \Omega -i \gamma  g_0)^2}\right)$$
of the equation
\begin{eqnarray}
(z+\gamma )(z -i \Omega ) -i g_0 \gamma  z=0,
\end{eqnarray}
one can rewrite the equation for the creation operator in the following form
\begin{eqnarray}
A^{+}(z)&=&a^{+}(0)\frac{z+\gamma -i g_{0} \gamma }{(z+\gamma )(z -i \Omega ) -i g_{0} \gamma  z}+F^{+}(z)\frac{i (z+\gamma )}{(z+\gamma )(z -i \Omega ) -i g_{0} \gamma  z}\\
&=&a^{+}(0)\frac{z+\gamma -i g_0 \gamma }{(z-z_1)(z-z_2)}+F^{+}(z)\frac{i (z+\gamma )}{(z-z_1)(z-z_2)}.
\label{eq_apl2}
\end{eqnarray}
The exact solution  $a^+(t)$  in terms of roots $z_i$ can be given by the residue theorem.
Then, the  explicit solution  for the original  is
\begin{eqnarray}
a^+(t)=a^+(0)\tilde{A}^*(t)+i \tilde{F}^+(t),
\label{eq_aaa}
\end{eqnarray}
where
\begin{eqnarray}
\tilde{A}^*(t)&=&L^{-1}\left\{ \frac{z+\gamma -i g_0 \gamma }{(z-z_1)(z-z_2)} \right\}\\ \nonumber
&=&
e^{t z_1}\frac{(z_1+ \gamma -i g_0 \gamma )}{z_1-z_2}+e^{t z_2}\frac{(z_2+ \gamma -i g_0 \gamma )}{z_2-z_1}
\end{eqnarray}
and
\begin{eqnarray}
\tilde{F}^+(t)&=&L^{-1}\left\{ F^{+}(z)\frac{i (z+\gamma )}{(z-z_1)(z-z_2)} \right\}\\ \nonumber
&=&\sum _{\nu } \frac{g_{\nu }}{\hbar }{\tilde{a}_{\nu }}^+(0)\left[\frac{e^{t z_1} (z_1+\gamma )}{(z_1-z_2) \left(z_1-i \omega _{\nu }\right)}+\frac{e^{t z_2} (-z_2-\gamma )}{(z_1-z_2) \left(z_2-i \omega _{\nu }\right)}+\frac{e^{i t \omega _{\nu }} \left(i \gamma -\omega _{\nu }\right)}{\left(z_2-i \omega _{\nu }\right) \left(i z_1+\omega _{\nu }\right)}\right].
\end{eqnarray}
%One should keep in mind that $\tilde{F}^+(t)\sim \tilde{a}_{\nu}^{+}(0)$.
Here, $L^{-1}$ is the inverse Laplace transform.

\section*{Appendix B}

The constants in the Eq. (23) are the following:
$$f_1=\frac{w^2+\gamma ^2}{\left(w^2-2 g_0 w \gamma +\left(1+g_0^2\right) \gamma ^2\right) \left(w^2-2 w \Omega +\left(1+g_0^2\right) \Omega ^2\right)},$$
$$f_2=\frac{g_0^2 \gamma ^2}{\left(w^2-2 g_0 w \gamma +\left(1+g_0^2\right) \gamma ^2\right) \left(\left(1+g_0^2\right) \gamma ^2-4 g_0 \gamma  \Omega +\left(1+g_0^2\right) \Omega ^2\right)},$$
$$f_3=\frac{\gamma ^2-2 g_0 \gamma  \Omega +\left(1+g_0^2\right) \Omega ^2}{\left(w^2-2 w \Omega +\left(1+g_0^2\right) \Omega ^2\right) \left(\left(1+g_0^2\right) \gamma ^2-4 g_0 \gamma  \Omega +\left(1+g_0^2\right) \Omega ^2\right)},$$
$$f_4=\frac{i g_0 (w+i \gamma ) \gamma }{\left(w^2-2 g_0 w \gamma +\left(1+g_0^2\right) \gamma ^2\right) (\gamma -i g_0 \gamma -(-i+g_0) \Omega ) (w+i (i+g_0) \Omega )},$$
$$f_5=\frac{(-i w+\gamma ) (\gamma -(-i+g_0) \Omega )}{(w-(-i+g_0) \gamma ) ((i+g_0) \gamma +(-1-i g_0) \Omega ) \left(w^2-2 w \Omega +\left(1+g_0^2\right) \Omega ^2\right)},$$
$$f_6=\frac{g_0 \gamma  (\gamma -(i+g_0) \Omega )}{(i w+\gamma -i g_0 \gamma ) (w+i (i+g_0) \Omega ) \left(\left(1+g_0^2\right) \gamma ^2-4 g_0 \gamma  \Omega +\left(1+g_0^2\right) \Omega ^2\right)}.$$
One can show that at $t=0$
$$f_1+f_2+f_3+f_4+f_4^{*}+f_5+f_5^{*}+f_6+f_6^{*}=0.$$

%\newpage

%\begin{figure}[h]
%\includegraphics[scale=1.0]{figure2.pdf}
%\caption{(Color online) The dependence of the  average  occupation number $n(t)$ on time $t$ for the fermionic (the notation is "f") and
%bosonic (the notation is "f") systems. The calculations for different coupling constants $g_0$ and temperatures $T$ are indicated on the figure. Here,
%$n(t=0)$=1.}
%\label{n_time_decay}
%\end{figure}

\end{document}